\documentclass[preprint2]{emulateapj}

\newcommand{\bzcat}{ROMA-BZCAT}

\newcommand{\fer}{{\it Fermi}}

\newcommand{\swf}{{\it Swift}}

\newcommand{\wse}{{\it WISE}}

\usepackage{graphicx}
\usepackage{longtable}

\slugcomment{version \today: fm}
\shorttitle{Unidentified Gamma-ray Sources V}
\shortauthors{F. Massaro et al. 2013}

\begin{document}
\title{Unveiling the nature of the unidentified gamma-ray sources V: \\ analysis of the radio candidates with the kernel density estimation}
\author{
F. Massaro\altaffilmark{1}, 
R. D'Abrusco\altaffilmark{2}, 
A. Paggi\altaffilmark{2}, 
N. Masetti\altaffilmark{3},\\
M. Giroletti\altaffilmark{4},
G. Tosti\altaffilmark{5},
Howard, A. Smith\altaffilmark{2},
\& 
S. Funk\altaffilmark{1}.
}

\altaffiltext{1}{SLAC National Laboratory and Kavli Institute for Particle Astrophysics and Cosmology, 2575 Sand Hill Road, Menlo Park, CA 94025, USA}
\altaffiltext{2}{Harvard - Smithsonian Astrophysical Observatory, 60 Garden Street, Cambridge, MA 02138, USA}
\altaffiltext{3}{INAF - Istituto di Astrofisica Spaziale e Fisica Cosmica di Bologna, via Gobetti 101, 40129, Bologna, Italy}
\altaffiltext{4}{INAF Istituto di Radioastronomia, via Gobetti 101, 40129, Bologna, Italy}
\altaffiltext{5}{Dipartimento di Fisica, Universit\`a degli Studi di Perugia, 06123 Perugia, Italy}

\begin{abstract}
Nearly one-third of the $\gamma$-ray sources detected by \fer\ are still unidentified, despite significant recent progress in this effort.
On the other hand, all the $\gamma$-ray extragalactic sources associated in the second \fer-LAT catalog have a radio counterpart.
Motivated by this observational evidence we investigate all the radio sources of the major radio surveys
that lie within the positional uncertainty region of the unidentified $\gamma$-ray sources (UGSs) at 95\% level of confidence. 
First we search for their infrared counterparts in the all-sky survey performed by the  
Wide-field Infrared Survey Explorer (\wse) and then we analyze their IR colors in comparison with those of the known $\gamma$-ray blazars.
We propose a new approach, based on a 2-dimensional kernel density estimation (KDE) technique 
in the single [3.4]-[4.6]-[12] $\mu$m \wse\ color-color plot,
replacing the constraint imposed in our previous investigations on the detection at 22$\mu$m 
of each potential IR counterpart of the UGSs with associated radio emission.
The main goal of this analysis is to find distant $\gamma$-ray blazar 
candidates that, being too faint at 22$\mu$m, are not detected by \wse\ and thus are not selected by our purely IR based methods.
We find fifty-five UGS's likely correspond to radio sources with blazar-like IR signatures.
Additional eleven UGSs having, blazar-like IR colors, have been found within the sample of sources
found with deep recent ATCA observations.
\end{abstract}


\section{Introduction}
\label{sec:intro}
The large majority of the point sources detected by the Compton Gamma-ray Observatory in the 1990s \citep[e.g.,][]{hartman99}
are still lacking an association with a low-energy candidate counterpart, and given their sky distribution,
a significant fraction of these unresolved objects are expected to have extragalactic origin \citep[e.g.,][]{thompson08,abdo10a}. 
Unveiling the origin of the unidentified $\gamma$-ray sources (UGSs) is also one of the
key scientific objectives of the recent \fer\ mission that still lists about 1/3 of the $\gamma$-ray sources as unassociated
in the second \fer-LAT catalog \citep[2FGL;][]{nolan12} .

A large fraction of UGSs is expected to be blazars, the largest known population of $\gamma$-ray active galaxies, 
not yet associated and/or recognized due to the lack of multifrequency observations \citep{ackermann11a}.
Therefore a better understanding of the nature of the UGSs is crucial to estimate accurately 
the blazar contribution to the extragalactic gamma-ray background \citep[e.g., ][]{mukherjee97,abdo10b},
and it is essential to constrain exotic high-energy physics phenomena \citep[e.g.,][]{zechlin12}.

Many attempts have been adopted to decrease UGSs number and to understand their composition. 
Pointed \swf\ observations\\
\citep[e.g.,][]{mirabal09a,mirabal09b,ugs4} to search for X-ray counterparts of UGSs
as well as radio follow up observations were already performed or are still in progress \citep[e.g.,][]{kovalev09a,kovalev09b,petrov13}.
In addition, statistical approaches based on different techniques have been also developed and successfully used \citep[e.g.][]{mirabal10,ackermann12}.
 
We recently addressed the problem of searching $\gamma$-ray blazar candidates 
as counterparts of the UGSs adopting two new approaches:
the first is based on the Wide-field Infrared Survey Explorer (\wse) all-sky observations \citep{wright10}
aiming at recognizing $\gamma$-ray blazar candidates using their peculiar IR colors 
\citep{paper1,paper2,paper4,ugs1} while the second employs the low-frequency radio observations \citep{ugs3}.
In particular, this second method was indeed based on the combination of the 
radio observations Westerbork Northern Sky Survey \citep[WENSS;][]{rengelink97}
at 325 MHz with those of the NRAO Very Large Array Sky survey
\citep[NVSS;][]{condon98} and of the Very Large Array Faint Images of the Radio Sky 
at Twenty-Centimeters \citep[FIRST;][]{becker95,white97} at about 1.4 GHz.
 
It is worth noting that all the \fer\ extragalactic sources associated in the 2FGL catalog have a clear radio counterpart \citep{nolan12},
this is the basis of the radio-$\gamma$-ray connection,
that has been found in the case of blazars \citep[e.g.,][]{ghirlanda10,mahony10,ackermann11b}.
Thus, motivated by this observational evidence we propose a different approach to search for the 
blazar-like counterparts of the UGSs.
We combine the radio and the IR information available for the sources lying within the positional uncertainty regions 
of the \fer\ UGSs to select $\gamma$-ray blazar candidates.

With respect to our previous IR based search for blazar-like counterparts\\
\citep[e.g.,][]{paper3,ugs1}
our new analysis relaxes the constraint on the 22$\mu$m detection of
the \wse-selected candidates, and does not take into account their [12]-[22] $\mu$m color, 
replacing these features with the presence of a radio counterpart.
The number of $\gamma$-ray blazars undetected at 22$\mu$m is only a small fraction \citep[$\sim$8\%of the total number of $\gamma$-ray blazars][]{ugs1},
but includes several high redshift sources that lying at larger distance than the whole population.

To perform our analysis, we search all the radio sources detected in the \\
NVSS \citep{condon98} and in the Sydney University Molonglo Sky Survey \citep[SUMSS;][]{mauch03} 
surveys that lie within the positional uncertainty region,
at 95\% level of confidence, of the UGSs listed in the 2FGL.
Then we associate them with their \wse\ counterparts to compare their IR colors with those of the known $\gamma$-ray blazars
in the [3.4]-[4.6]-[12] $\mu$m plot using the kernel density estimation (KDE) technique \citep[e.g.,][]{richards04,dabrusco09,paper3}.
We also verified if the radio sources found in the recent deep radio observations performed by 
Australia Telescope Compact Array (ATCA) and presented by Petrov et al. (2013)
have an IR counterpart with \wse\ colors consistent with those of the $\gamma$-ray blazar population.
Our analysis of the IR colors is restricted only to the [3.4]-[4.6]-[12] $\mu$m color-color plot.

The paper is organized as follows: 
Section~\ref{sec:sample} is devoted to the definitions of the samples used while
in Section~\ref{sec:kde} we describe the KDE technique used to perform our investigation;
we then applied our selection in Section~\ref{sec:ugs} to identify those 
radio sources that could be considered blazar-like counterpart of the UGSs listed in the 2FGL catalog.
We also verified the presence of optical and X-ray counterparts for the selected $\gamma$-ray blazar candidates and
we compare our results with different approaches previously developed.
Finally, Section~\ref{sec:summary} is dedicated to our conclusions. 

For our numerical results, we use cgs units unless stated otherwise.
Spectral indices, $\alpha$, are defined by flux density, S$_{\nu}\propto\nu^{-\alpha}$ and
\wse\ magnitudes at the [3.4], [4.6], [12], [22] $\mu$m (i.e., the nominal \wse\ bands)
are in the Vega system respectively.
All the magnitudes and the IR colors reported in the paper have been corrected for the Galactic extinction
according to the formulae reported in Draine (2003) as also performed in our previous analysis \citep[e.g.,][]{ugs1,ugs2}.
The most frequent acronyms used in the paper are listed in Table~\ref{tab:acronym}.
\begin{table}
\caption{List of most frequent acronyms.}
\begin{tabular}{|lc|}
\hline
Name & Acronym \\
\hline
\noalign{\smallskip}
Multifrequency Catalog of blazars & \bzcat\ \\ 
Second \fer\ Large Area Telescope Catalog & 2FGL \\
\hline
\noalign{\smallskip}
BL Lac object & BZB \\
Flat Spectrum Radio Quasar & BZQ \\
Blazar of Uncertain type & BZU \\
Unidentified Gamma-ray Source & UGS \\
\hline
\noalign{\smallskip}
Training Blazar sample & TB \\
Northern UGS sample & NU \\
Southern UGS sample & SU \\
Southern Deep ATCA sample & SDA \\ 
\hline
\noalign{\smallskip}
Kernel Density Estimation & KDE \\
\noalign{\smallskip}
\hline
\end{tabular}\\
\label{tab:acronym}
\end{table}

\section{Sample selection}
\label{sec:sample}
The first sample used in our analysis lists all the blazars listed in the Multiwavelength Blazar 
Catalog\footnote{http://www.asdc.asi.it/bzcat/} \citep[\bzcat,][]{massaro09} that have been associated
as counterparts of \fer\ sources in the 2FGL \citep{nolan12} with a \wse\ counterpart detected at least in the first three filters
regardless of the fact that they are detected at 22$\mu$m. The association radius between the \bzcat\ catalog and the \wse\ all-sky survey adopted here 
was fixed to 3\arcsec.3 \citep[see][for more details]{ugs1}.
This sample, named {\it training blazar} (TB) sample, comprises a total of 737 blazars, 
excluding those classified as blazars of uncertain type (BZUs) \citep[see also][]{massaro10,massaro11}.
The TB sample is used to build the isodensity contours for the KDE technique (see following sections) and to test
if IR sources with radio counterparts have \wse\ colors consistent with the $\gamma$-ray blazar population.

Then the UGSs sample considered is the one constituted by all the \fer\ sources 
listed in the 2FGL with no assigned counterpart at low energies and
without any $\gamma$-ray analysis flag listing 299 sources \citep{nolan12}.
We further divided this sample in two subsamples: the northern UGS (NU) sample where only sources
with Declination above than -40 deg and the southern UGS (SU) sample selecting those at Declination below -30 deg.
This subdivision has been chosen on the basis of the footprints of the radio surveys used for our analysis,
since the NU sample is mainly covered by the NVSS survey \citep{condon98},
while the SU one by the SUMSS catalog \citep{mauch03}.
The former sample lists 209 UGSs while 115 sources belong to the latter one.

Finally, we also considered the list of all the radio sources recently found by Petrov et al. (2013) 
using deep ATCA observations for the UGSs in the southern hemisphere. 
This sample is labeled as southern deep ATCA (SDA) sample.

\section{Kernel Density Estimation}
\label{sec:kde}
The KDE technique is a non-parametric procedure to estimate the probability density function 
of a multivariate distribution without requiring any assumption about the shape of the ``parent'' distribution.
The KDE technique also permits to reconstruct the density distribution of a population 
of points in a general N-dimensional space based on a finite sample. 
This analysis depends on only one parameter, the bandwidth of the kernel of the density estimator 
(analogous to the window size for one-dimensional running average) that can be estimated locally
\citep[see e.g., ][and reference therein]{richards04,dabrusco09,laurino11}.

We already applied the KDE technique in several cases to compare the IR colors of blazar candidates selected with different procedures
with those of the known population of $\gamma$-ray blazars \citep[see][for more details]{paper1,paper3,ugs4}.
Thus in the present analysis we use the KDE method to compare the IR colors of the radio selected 
counterparts with those of the $\gamma$-ray blazar population represented by the TB sample
in the 2-dimensional [3.4]-[4.6]-[12] $\mu$m color-color plot.
As already described in Massaro et al. (2012a), we provide an associated confidence $\pi_{kde}$
drawn from the KDE density probabilities that a selected radio source as IR colors consistent with the blazars in the TB sample.

\begin{figure*}[] 
           \includegraphics[height=7.8cm,width=8.8cm,angle=0]{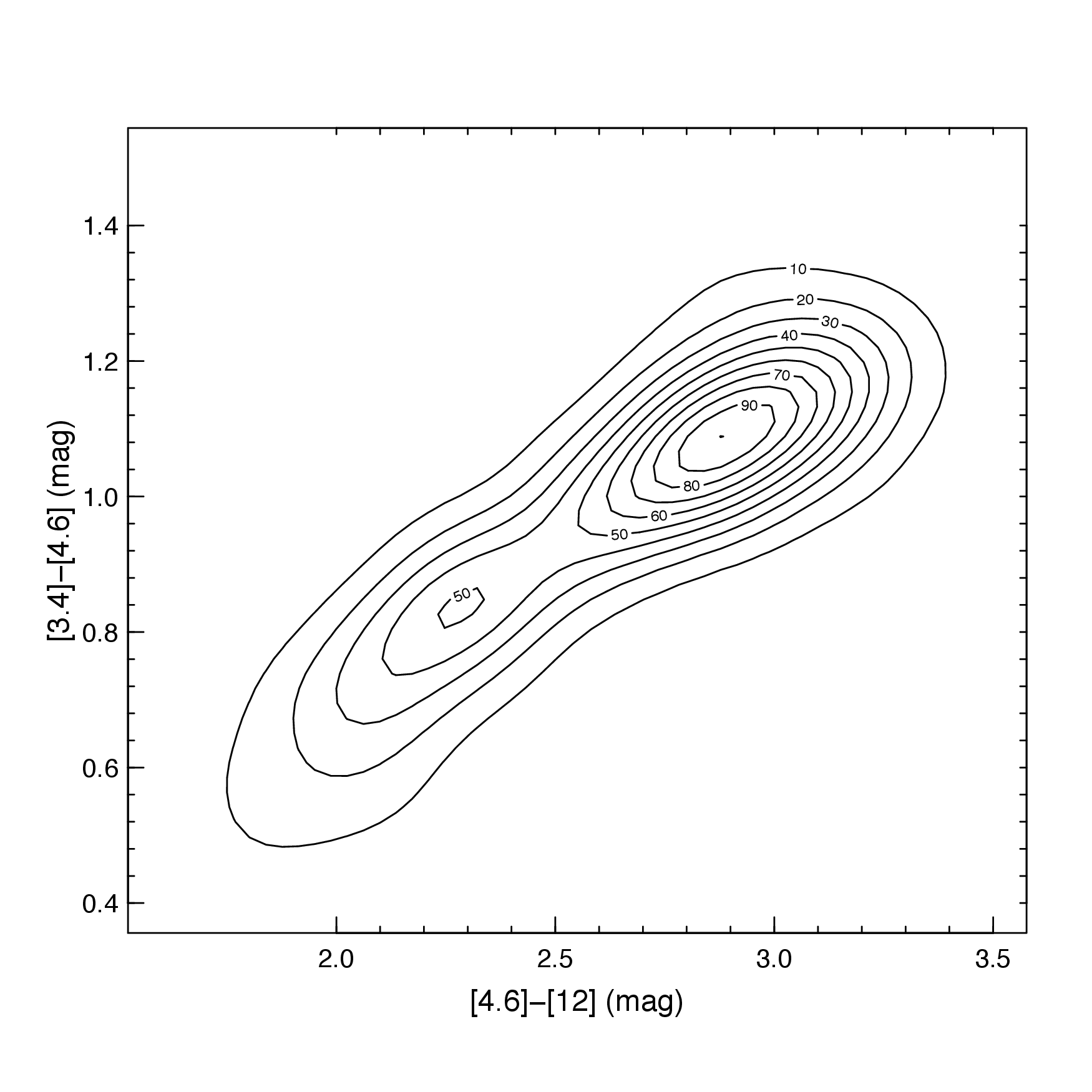}
           \includegraphics[height=7.8cm,width=8.8cm,angle=0]{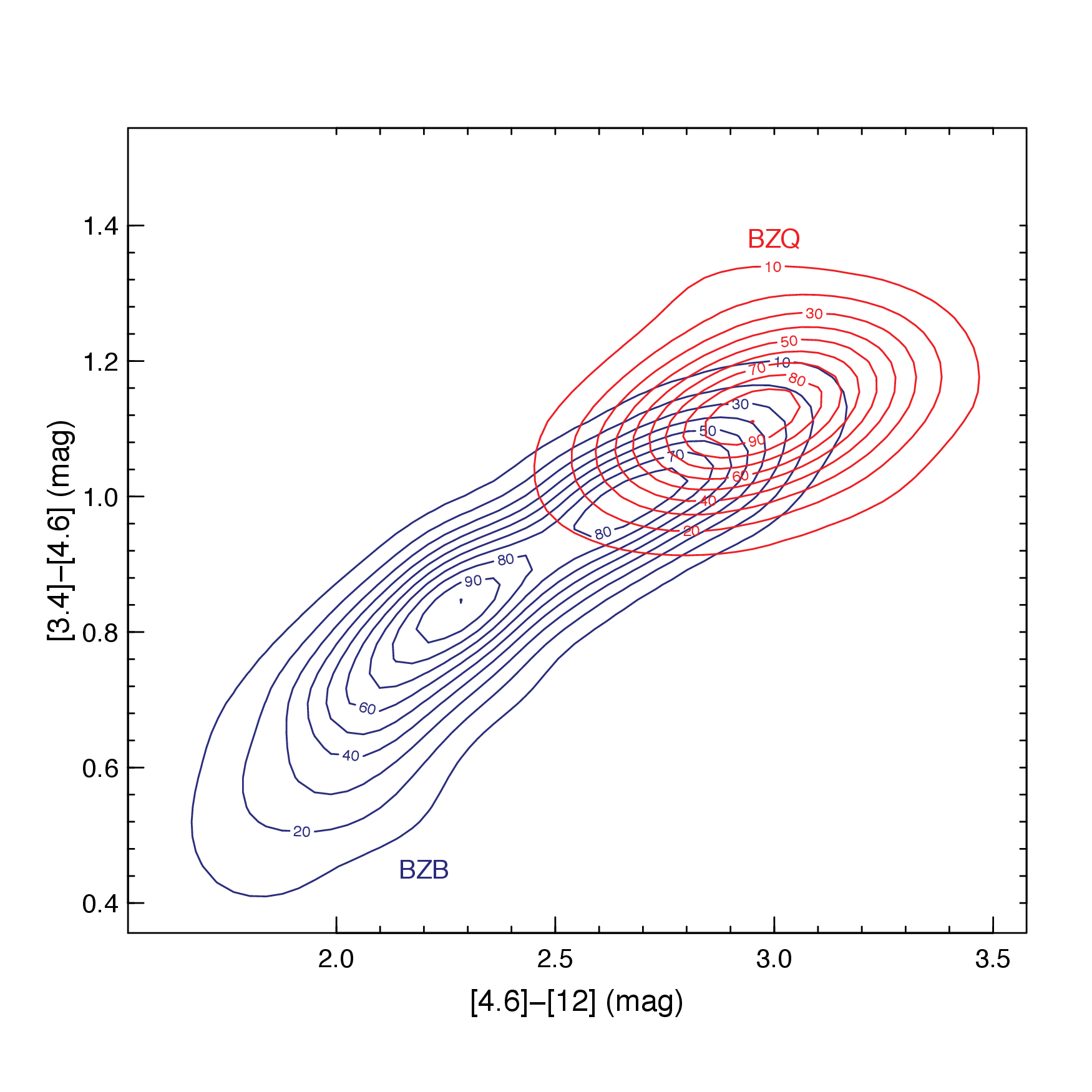}
           \caption{Left) The isodensity contours generated by KDE technique in the
                                 [3.4]-[4.6]-[12] $\mu$m color-color diagram for the whole $\gamma$-ray
                                 blazar population represented by the sources in the TB sample.
                        Right) The KDE isodensity contours built separately for the BZB (blue) and the BZQ (red) classes in the TB sample.
                        The numbers appearing close to each contour corresponds to the values of $\pi_{kde}$ in both panels.}
          \label{fig:kde}
          \end{figure*}

In Figure~\ref{fig:kde} we show the density profiles constructed for the whole blazar population (left panel)
and used to estimate $\pi_{kde}$ and those of the two subsamples of BZBs and BZQs (right panel) belonging to the TB sample,
to highlight the dichotomy between the two subclasses.  

\section{Unidentified $\gamma$-ray Sources}
\label{sec:ugs}

\subsection{Selection of $\gamma$-ray blazar candidates}
\label{sec:candidates}
For each UGS we searched for all the radio sources that lie within their positional uncertainty regions at 95\%
level of confidence and we found that there are 822 radio sources potential counterparts of 209 UGSs and
134 out of 115 for the NU and the SU samples, respectively. 
We then crossmatched all these radio sources with the \wse\ all-sky catalog\footnote{http://wise2.ipac.caltech.edu/docs/release/allsky/} \citep{wright10}
using the same radius of 3\arcsec.3 and we selected only 
those with an IR counterpart detected at least in the first three \wse\ filters and not extended (i.e., extension flag, $ext\_flg\leq$ 1) \citep{cutri12}.
The 3\arcsec.3 radius chosen to associated sources between the \wse\ and the radio catalogs 
is statistically justified on the basis of the analysis performed over the entire \bzcat\ \citep[see][for more details]{ugs1}.
Thus we obtained 374 out of 822 and 78 out of 134  radio sources in the NU and SU samples, respectively.

Subsequently, we applied the KDE technique described in Section~\ref{sec:kde} to find radio sources with \wse\ counterparts
having IR colors consistent with the $\gamma$-ray blazar population.
We considered reliable $\gamma$-ray blazar candidates only radio sources consistent 
within the isodensity contours, drawn from the KDE, at 90\% level of confidence,
correspondent to an association confidence ($\pi_{kde}$) grater than 10.0.

We found 41 and 14 radio sources \wse\ selected with $\pi_{kde}>$0.1 within the NU and the SU samples, respectively.
In addition, only 11 out of 416 radio sources listed in the SDA sample have an IR counterpart 
consistent with the \fer\ blazar population of the TB sample with $\pi_{kde}>$10.0.
We also list two exceptions to the above criteria: the UGS 2FGLJ1223.3+7954 with its \wse\ blazar candidate WISE J122358.17+795327.8
in the NU sample and 2FGLJ0523.3-2530 with WISE J052313.07-253154.4 as potential counterpart
in the SDA sample, having the $\pi_{kde}$ values equal to 9.6 and 9.5, respectively, marginally below our threshold.
The total number of $\gamma$-ray blazar candidates is 66 all listed in Table~\ref{tab:table1} and Table~\ref{tab:table2}.
It is worth noting that we do not have any multiple $\gamma$-ray blazar candidate within
the positional uncertainty regions of the UGSs analyzed.

\begin{figure}[!b] 
           \includegraphics[height=7.8cm,width=8.8cm,angle=0]{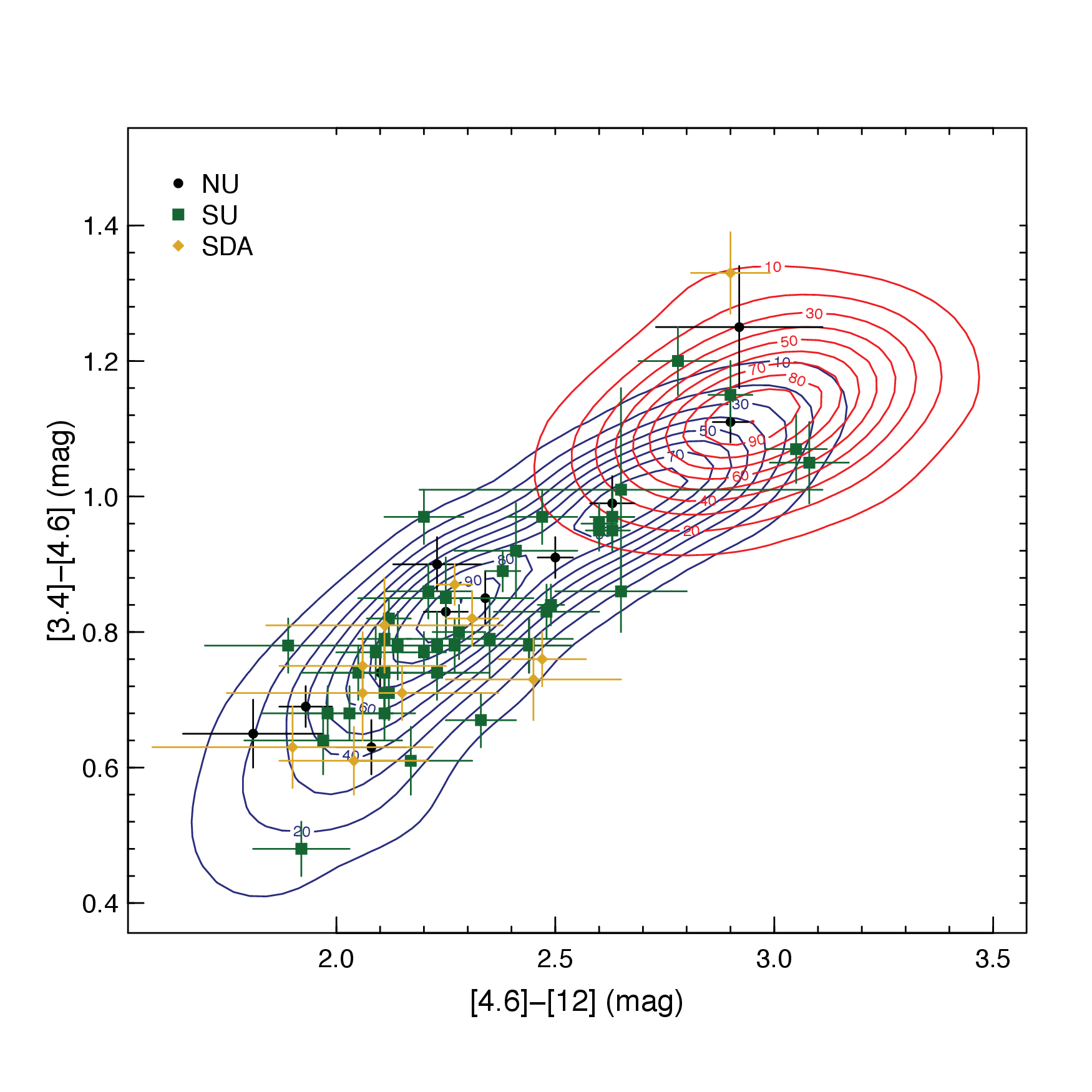}
           \caption{The isodensity contours generated by KDE technique in the
                         [3.4]-[4.6]-[12] $\mu$m color-color diagram for the BZBs (blue) and the BZQs (red)
                         in the TB sample. Points overlaid to the contours show the location of the selected radio candidates
                         with IR colors consistent with the $\gamma$-ray blazar population within $\pi_{kde}>$10
                         for the sources in the three different samples analyzed: NU (black circles), SU (green squares) and SDA (yellow diamonds).
                         The numbers appearing close to each contour corresponds to the values of $\pi_{kde}$.}
          \label{fig:ugs}
          \end{figure}

In Figure~\ref{fig:ugs} we show the isodensity contours derived from the KDE analysis in the [3.4]-[4.6]-[12] $\mu$m color color plot,
together with the $\gamma$-ray blazar candidates selected in the UGS samples analyzed and in the SDA list.
It is evident how the large fraction for the selected candidates are located within with the isodensity contours drawn for the BZB class.

To establish if the $\gamma$-ray blazar candidate selected with our method have additional multifrequency properties 
that could confirm their nature and provide redshift estimates, we also searched for the counterpart 
of our radio-IR selected candidates in the following major surveys.
For the near-IR we used only the Two Micron All Sky Survey \citep[2MASS;][- M]{skrutskie06}
since each \wse\ source is already associated with the closest 2MASS source 
by the default catalog \citep[see][for more details]{cutri12}.
We then searched for optical counterparts, with possible spectra available, 
in the Sloan Digital Sky Survey \citep[SDSS; e.g.][- s]{adelman08,paris12}, in the Six-degree-Field Galaxy Redshift Survey 
\citep[6dFGS;][- 6]{jones04,jones09}, in the The Muenster Red Sky Survey \citep[MRSS;][]{ungruhe03} and in the USNO-B Catalog \citep{monet03} 
within 3\arcsec.3. These optical cross correlations are also useful to plan 
follow up observations thus a complete list of sources together with their optical magnitudes is reported in Table~\ref{tab:optical}.
For the high energy we looked in the soft X-rays using the ROSAT all-sky survey catalog \citep[RASS;][- X]{voges99}.
Finally, we also considered the NASA Extragalactic Database (NED)
\footnote{\underline{http://ned.ipac.caltech.edu/}} for any possible counterpart within 3\arcsec.3 for additional information.
The results of this multifrequency investigation is presented and summarized in Table~\ref{tab:table1} and Table~\ref{tab:table2}.
\begin{table*}
\tiny
\caption{Unidentified Gamma-ray Sources in the Northern and in the Southern samples.}
\begin{tabular}{|lllccclcl|}
\hline
  2FGL  &  WISE      &  Radio  & [3.4]-[4.6] & [4.6]-[12] & $\pi_{kde}$ & notes & z  & compare \\
  name  &  name      &  name  &     mag     &    mag     &             &       &    &         \\
\hline
\noalign{\smallskip}
NORTHERN UGS SAMPLE & & & & & & & & \\
\noalign{\smallskip}
\hline
  2FGLJ0031.0+0724  &  J003119.70+072453.6  &  NVSSJ003119+072456  &  0.83(0.04)  &  2.48(0.12)  &  29.3  &  N           &  ?  &  3        \\
  2FGLJ0039.1+4331  &  J003908.14+433014.6  &  NVSSJ003907+433015  &  0.97(0.04)  &  2.20(0.09)  &  10.3  &  N,v         &  ?  &  1,2,3    \\
  2FGLJ0103.8+1324  &  J010345.73+132345.4  &  NVSSJ010345+132346  &  0.68(0.04)  &  2.03(0.10)  &  31.3  &  N,M         &  ?  &  3        \\
  2FGLJ0158.4+0107  &  J015852.76+010132.9  &  NVSSJ015852+010133  &  0.85(0.06)  &  2.25(0.20)  &  49.1  &  N,F,s,rv    &  ?  &  -        \\ 
  2FGLJ0158.6+8558  &  J015248.80+855703.6  &  NVSSJ015248+855706  &  1.07(0.05)  &  3.05(0.07)  &  65.6  &  N,M         &  ?  &  1,2      \\
  2FGLJ0227.7+2249  &  J022744.35+224834.3  &  NVSSJ022744+224834  &  0.95(0.03)  &  2.60(0.03)  &  53.1  &  N,v       &  ?  &  1!,3!    \\
  2FGLJ0312.8+2013  &  J031240.54+201142.8  &  NVSSJ031240+201141  &  0.79(0.06)  &  2.35(0.19)  &  36.4  &  N           &  ?  &  -        \\
  2FGLJ0332.1+6309  &  J033153.90+630814.1  &  NVSSJ033153+630814  &  0.96(0.03)  &  2.60(0.04)  &  54.5  &  N,M         &  ?  &  1!,2!    \\
  2FGLJ0353.2+5653  &  J035309.54+565430.8  &  NVSSJ035309+565431  &  0.78(0.04)  &  1.89(0.19)  &  10.9  &  N,M,rv      &  ?  &  2!,3!    \\
  2FGLJ0409.8-0357  &  J040946.57-040003.4  &  NVSSJ040946-040003  &  0.89(0.03)  &  2.38(0.04)  &  46.4  &  N,M         &  ?  &  1!,3!    \\
  2FGLJ0420.9-3743  &  J042025.09-374445.0  &  NVSSJ042025-374443  &  0.78(0.04)  &  2.44(0.10)  &  20.2  &  N,S         &  ?  &  3!       \\
  2FGLJ0600.9+3839  &  J060102.86+383829.2  &  NVSSJ060102+383828  &  0.97(0.04)  &  2.47(0.08)  &  38.3  &  N           &  ?  &  2!,3!    \\
  2FGLJ0644.6+6034  &  J064435.72+603851.2  &  NVSSJ064435+603849  &  0.64(0.05)  &  1.97(0.18)  &  24.6  &  N           &  ?  &  1,2!,3   \\
  2FGLJ0658.4+0633  &  J065845.02+063711.5  &  NVSSJ065844+063711  &  0.68(0.04)  &  1.98(0.15)  &  27.4  &  N           &  ?  &  3        \\
  2FGLJ0723.9+2901  &  J072354.83+285929.9  &  NVSSJ072354+285930  &  1.15(0.05)  &  2.90(0.05)  &  81.0  &  N,F         &  ?  &  1!,2!,3! \\
  2FGLJ0746.0-0222  &  J074627.03-022549.3  &  NVSSJ074627-022549  &  0.68(0.04)  &  2.11(0.07)  &  31.3  &  N,M         &  ?  &  1!,3!    \\
  2FGLJ0928.8-3530  &  J092849.83-352948.9  &  NVSSJ092849-352947  &  0.97(0.04)  &  2.63(0.05)  &  57.8  &  N,S,M       &  ?  &  -        \\
  2FGLJ1016.1+5600  &  J101544.44+555100.7  &  NVSSJ101544+555100  &  1.05(0.06)  &  3.08(0.09)  &  48.0  &  N,F,s       &  ?  &  1!,2!    \\ 
  2FGLJ1115.0-0701  &  J111511.74-070239.9  &  NVSSJ111511-070238  &  0.86(0.06)  &  2.65(0.15)  &  17.2  &  N           &  ?  &  3        \\
  2FGLJ1123.3-2527  &  J112325.38-252857.0  &  NVSSJ112325-252858  &  0.84(0.03)  &  2.49(0.03)  &  30.0  &  N,M,6,QSR  &  0.146  &  -        \\
  2FGLJ1129.5+3758  &  J112903.25+375657.4  &  NVSSJ112903+375655  &  0.92(0.07)  &  2.41(0.14)  &  42.3  &  N,F,M,s,BL?     &  ?  &  3        \\ 
  2FGLJ1223.3+7954  &  J122358.17+795327.8  &  NVSSJ122358+795329  &  0.48(0.04)  &  1.92(0.11)  &   9.6  &  N,M         &  ?  &  2!,3     \\
  2FGLJ1254.2-2203  &  J125422.47-220413.6  &  NVSSJ125422-220413  &  0.67(0.04)  &  2.33(0.08)  &  11.4  &  N,M,v     &  ?  &  1!,3!    \\
  2FGLJ1259.8-3749  &  J125949.80-374858.1  &  NVSSJ125949-374856  &  0.71(0.04)  &  2.11(0.08)  &  36.8  &  N,S,M,v     &  ?  &  1!,3!    \\
  2FGLJ1340.5-0412  &  J134042.02-041006.8  &  NVSSJ134042-041006  &  0.71(0.04)  &  2.12(0.08)  &  36.6  &  N,M,v       &  ?  &  1!       \\
  2FGLJ1347.0-2956  &  J134706.89-295842.3  &  NVSSJ134706-295840  &  0.79(0.03)  &  2.11(0.06)  &  39.8  &  N,S,M,v     &  ?  &  1!,3!    \\
  2FGLJ1513.5-2546  &  J151303.66-253925.9  &  NVSSJ151303-253924  &  1.01(0.15)  &  2.65(0.46)  &  65.9  &  N           &  ?  &  3        \\
  2FGLJ1517.2+3645  &  J151649.26+365022.9  &  NVSSJ151649+365023  &  0.95(0.03)  &  2.63(0.04)  &  54.5  &  N,F,s,v     &  ?  &  1!,2,3   \\ 
  2FGLJ1548.3+1453  &  J154824.39+145702.8  &  NVSSJ154824+145702  &  0.74(0.05)  &  2.11(0.19)  &  39.6  &  N,F,M,s     &  ?  &  -        \\ 
  2FGLJ1647.0+4351  &  J164619.95+435631.0  &  NVSSJ164619+435631  &  0.77(0.04)  &  2.09(0.09)  &  38.1  &  N,F,s,X     &  ?  &  1!       \\
  2FGLJ1704.3+1235  &  J170409.59+123421.7  &  NVSSJ170409+123421  &  0.74(0.04)  &  2.05(0.07)  &  35.4  &  N,M         &  ?  &  3        \\
  2FGLJ1704.6-0529  &  J170433.84-052840.6  &  NVSSJ170433-052839  &  0.78(0.05)  &  2.14(0.16)  &  43.0  &  N,M,v       &  ?  &  3        \\
  2FGLJ2004.6+7004  &  J200506.02+700439.3  &  NVSSJ200506+700440  &  0.77(0.03)  &  2.20(0.05)  &  45.7  &  N,v         &  ?  &  1!,3     \\
  2FGLJ2021.5+0632  &  J202155.45+062913.7  &  NVSSJ202155+062914  &  0.82(0.03)  &  2.12(0.05)  &  35.3  &  N,M         &  ?  &  1!,3!    \\
  2FGLJ2115.4+1213  &  J211522.00+121802.8  &  NVSSJ211522+121802  &  0.78(0.05)  &  2.23(0.18)  &  46.2  &  N,M         &  ?  &  3!       \\
  2FGLJ2132.5+2605  &  J213253.05+261143.8  &  NVSSJ213252+261143  &  1.20(0.05)  &  2.78(0.09)  &  25.9  &  N           &  ?  &  3        \\
  2FGLJ2133.9+6645  &  J213349.21+664704.3  &  NVSSJ213349+664706  &  0.80(0.04)  &  2.28(0.06)  &  49.0  &  N,v         &  ?  &  1!,2,3   \\
  2FGLJ2134.6-2130  &  J213430.18-213032.6  &  NVSSJ213430-213032  &  0.78(0.04)  &  2.27(0.08)  &  44.3  &  N,M         &  ?  &  1!,3     \\
  2FGLJ2228.6-1633  &  J222830.19-163642.8  &  NVSSJ222830-163643  &  0.74(0.04)  &  2.23(0.12)  &  37.9  &  N,M         &  ?  &  3!       \\
  2FGLJ2246.3+1549  &  J224604.98+154435.3  &  NVSSJ224604+154437  &  0.61(0.05)  &  2.17(0.14)  &  16.0  &  N,M         &  ?  &  3!       \\
  2FGLJ2358.4-1811  &  J235836.83-180717.3  &  NVSSJ235836-180718  &  0.86(0.04)  &  2.21(0.10)  &  43.2  &  N,M,6,X,BL  &  0.058?  &  1        \\
\hline
\noalign{\smallskip}
SOUTHERN UGS SAMPLE & & & & & & & & \\
\noalign{\smallskip}
\hline
  2FGLJ0116.6-6153  &  J011619.59-615343.5  &  SUMSSJ011619-615343 &  0.85(0.04)  &  2.34(0.06) &  49.9  &  S,M         &  ?  &  1!,3!    \\
  2FGLJ0133.4-4408  &  J013306.35-441421.3  &  SUMSSJ013306-441422 &  0.83(0.03)  &  2.25(0.05) &  51.0  &  S,M         &  ?  &  1!,3!    \\
  2FGLJ0143.6-5844  &  J014347.39-584551.3  &  SUMSSJ014347-584550 &  0.69(0.03)  &  1.93(0.06) &  23.0  &  S,M         &  ?  &  1!,3     \\
  2FGLJ0316.1-6434  &  J031614.31-643731.4  &  SUMSSJ031614-643732 &  0.74(0.03)  &  2.10(0.06) &  38.9  &  S,M         &  ?  &  1!,3     \\
  2FGLJ0416.0-4355  &  J041605.81-435514.6  &  SUMSSJ041605-435516 &  1.11(0.03)  &  2.90(0.04) &  97.2  &  S,M         &  ?  &  1!       \\
  2FGLJ0420.9-3743  &  J042025.09-374445.0  &  MRSS303-096250      &  0.78(0.04)  &  2.44(0.10) &  20.2  &  N,S         &  ?  &  3!       \\
  2FGLJ0555.9-4348  &  J055618.74-435146.1  &  SUMSSJ055618-435146 &  0.91(0.03)  &  2.50(0.04) &  43.9  &  S,M         &  ?  &  1!       \\
  2FGLJ0928.8-3530  &  J092849.83-352948.9  &  SUMSSJ092849-352947 &  0.97(0.04)  &  2.63(0.05) &  57.8  &  N,S,M       &  ?  &  -        \\
  2FGLJ1032.9-8401  &  J103015.35-840308.7  &  SUMSSJ103014-840307 &  0.99(0.04)  &  2.63(0.05) &  62.1  &  S,v         &  ?  &  1!       \\
  2FGLJ1259.8-3749  &  J125949.80-374858.1  &  SUMSSJ125949-374856 &  0.71(0.04)  &  2.11(0.08) &  36.8  &  N,S,M,v     &  ?  &  1!,3!    \\
  2FGLJ1328.5-4728  &  J132840.61-472749.2  &  SUMSSJ132840-472748 &  0.63(0.04)  &  2.08(0.08) &  24.4  &  S,M,v       &  ?  &  3!       \\
  2FGLJ2042.8-7317  &  J204201.92-731913.5  &  SUMSSJ204201-731911 &  0.65(0.05)  &  1.81(0.16) &  12.1  &  S,M         &  ?  &  -        \\
  2FGLJ2131.0-5417  &  J213208.28-542036.4  &  SUMSSJ213208-542037 &  1.25(0.09)  &  2.92(0.19) &  29.0  &  S           &  ?  &  -        \\
  2FGLJ2213.7-4754  &  J221330.33-475425.0  &  SUMSSJ221330-475426 &  0.90(0.04)  &  2.23(0.10) &  33.4  &  S,M         &  ?  &  -        \\
\noalign{\smallskip}
\hline
\end{tabular}\\
Col. (1) 2FGL name. \\
Col. (2) WISE name. \\
Col. (3) Radio name. \\
Cols. (4,5) IR colors from \wse. Values in parentheses are 1$\sigma$ uncertainties. \\
Col. (6) Notes: N = NVSS, F = FIRST, M = 2MASS, s = SDSS dr9, 6 = 6dFG; X=ROSAT; QSO  = quasar, BL = BL Lac; v = variable in \wse\ bands 
(var\_flag $>$ 5 in at least one band, see Cutri et al. 2012 for additional details); rv = variable in the radio bands at 1.4 GHz. \\
Col. (7) Estimate level of confidence derived from the KDE analysis.\\
Col. (8) Redshift: ? = unknown. \\
Col. (9) Results of the comparison with previous analyses. 1 = UGS analyzed in Massaro et al. (2013a) , 2 = UGS analyzed in Massaro et al. (2013b)
3 = UGS analyzed in Paggi et al. (2013). Exclamation mark (!) indicates that the $\gamma$-ray blazar candidate is the same IR source found
in the previous investigation.
\label{tab:table1}
\end{table*}

\subsection{Probability of spurious associations}
\label{sec:chance}
{ We estimated the probability that our $\gamma$-ray blazar candidates can be spurious associations
adopting the following approach, similar to that successfully used in our previous analyses \citep[e.g.,][]{ugs3,ugs4}.

We created two {\it fake} $\gamma$-ray catalogs
shifting the coordinates of the 41 $\gamma$-ray blazars in the NU sample 
and of the 25 in the SU one by 0\degr.7 in a random direction of the sky 
within the footprints of the NVSS and the SUMSS radio surveys.
Keeping the same values of $\theta_{95}$ of each {\it fake} UGS, 
we verified that there were no correspondences with real \fer\ sources  
within a circular region of radius $\theta_{95}$ at the flux level of the 2FGL.

For each {\it fake} UGSs, we search for all the radio sources lying 
within the positional uncertainty region 
at 95\% of confidence in both the NVSS and SUMSS radio surveys.
We then checked the presence of an IR counterpart of each radio source selected above
crossmatching the \wse\ all-sky catalog with their NVSS and SUMSS 
positions within a radius of 3\arcsec.3.
The value of this IR-to-radio association radius has been chosen on the basis of our previous
statistical analyses \citep[see Section~\ref{sec:sample} and ][for more details]{ugs1}.

For each radio source with a \wse\ counterpart we applied our KDE technique 
selecting the radio sources detected by \wse\ at 3.4$\mu$m, 4.5$\mu$m and 12$\mu$m
with $\pi_{kde}>0.10$ being {\it fake} $\gamma$-ray blazar candidates.
Then we repeated the entire procedure 10 times for both the NU and the SU sample
to establish the probability of spurious associations.
Based on the above procedure, we expect that 4\% and 3\% of the $\gamma$-ray blazar candidates
previously selected for the UGS in the NU and SU samples respectively, could be contaminants.

Finally, we emphasize that these estimates depend on the $\gamma$-ray background model,
the detection threshold and the flux limit of the 2FGL catalog \citep{nolan12},
in which no $\gamma$-ray emission is arising from 
any of the positions listed in the {\it fake} $\gamma$-ray catalogs.}

\subsection{Comparison with previous investigations}
\label{sec:comparison}
We compare our results with those of previous analyses carried out in 
Massaro et al. (2013a), Massaro et al. (2013b) and Paggi et al. (2013).
The results of our comparison is summarized below and presented in Table~\ref{tab:table1} and Table~\ref{tab:table2}.

We note that within the 41 $\gamma$-ray blazar candidates found in the NU sample
there are 16 sources that were also selected on the basis of their three \wse\ colors in Massaro et al. (2013a)
7 that appeared as potential counterpart in Massaro et al. (2013b) found with the low-frequency radio observations and 14
listed with an X-ray properties in Paggi et al. (2013). In addition, 12 UGS were also investigated in our previous analyses
but for them we found a different $\gamma$-ray blazar candidate.
The number of new candidates counterparts in the NU sample is 5.
On the other hand, within the SU sample, we found that 8 radio sources were also selected in Massaro et al. (2013a)
and 4 in Paggi et al. (2013), in addition to 4 new $\gamma$-ray blazar candidates.

Petrov et al. (2013) already found the \wse\ counterparts of their SDA sample but they 
did not verified which have IR colors consistent with the \fer\ blazars.
Thus in the SDA sample we listed 11 radio sources detected thanks to the deeper radio survey performed with ATCA \citep{petrov13}
with IR colors consistent with those of the $\gamma$-ray blazar population.
Among these 11 $\gamma$-ray blazar candidates, there are two sources already found in Massaro et al. (2013a)
and only one UGS (i.e., 2FGLJ0547.5-0141c) previously investigated that appear to have a different potential counterpart.
\begin{table*}
\tiny
\caption{Unidentified Gamma-ray sources in the SDA sample.}
\begin{tabular}{|lllccclcl|}
\hline
  2FGL  &  WISE      &  IAU   & [3.4]-[4.6] & [4.6]-[12] & $\pi_{kde}$ & notes & z  & compare \\
  name  &  name      &  name  &     mag     &    mag     &             &       &    &         \\
\hline
\noalign{\smallskip}
  2FGLJ0200.4-4105  &  J020020.94-410935.6  &  J0200-4109  &  0.63(0.06)  &  1.90(0.32)  &  19.3  & 6,X & ? & \\
  2FGLJ0340.7-2421  &  J034022.89-242407.2  &  J0340-2424  &  0.73(0.06)  &  2.45(0.20)  &  10.0  & N & ? & \\
  2FGLJ0523.3-2530  &  J052313.07-253154.4  &  J0523-2531  &  1.33(0.06)  &  2.90(0.09)  &   9.5  & - & ? & \\
  2FGLJ0547.5-0141c &  J054720.85-013329.9  &  J0547-0133  &  0.81(0.07)  &  2.11(0.27)  &  36.6  & N & ? & 1 \\
  2FGLJ0937.9-1434  &  J093754.72-143350.3  &  J0937-1433  &  0.71(0.04)  &  2.15(0.08)  &  35.1  & N & ? & \\
  2FGLJ1315.6-0730  &  J131552.98-073301.9  &  J1315-0733  &  0.87(0.03)  &  2.27(0.04)  &  47.2  & N,F,M,v,BL? & ? & 1! \\
  2FGLJ1339.2-2348  &  J133916.44-234829.4  &  J1339-2348  &  0.75(0.05)  &  2.06(0.19)  &  35.0  & N & ? & \\
  2FGLJ1345.8-3356  &  J134543.05-335643.3  &  J1345-3356  &  0.82(0.04)  &  2.31(0.06)  &  49.8  & N,S,M & ? & 1! \\
  2FGLJ2034.7-4201  &  J203451.08-420038.2  &  J2034-4200  &  0.61(0.05)  &  2.04(0.17)  &  22.3  & - & ? & \\
  2FGLJ2251.1-4927  &  J225128.69-492910.6  &  J2251-4929  &  0.76(0.04)  &  2.47(0.10)  &  12.1  & S & ? & \\
  2FGLJ2343.3-4752  &  J234302.29-475749.9  &  J2343-4757  &  0.71(0.07)  &  2.06(0.31)  &  35.4  & S & ? & \\
\noalign{\smallskip}
\hline
\end{tabular}\\
Col. (1) 2FGL name. \\
Col. (2) WISE name. \\
Col. (3) Radio name. \\
Cols. (4,5) IR colors from \wse. Values in parentheses are 1$\sigma$ uncertainties. \\
Col. (6) Notes: N = NVSS, F = FIRST, M = 2MASS, s = SDSS dr9, 6 = 6dFG;  X=ROSAT; QSO  = quasar, BL = BL Lac; v = variable in \wse\ bands 
(var\_flag $>$ 5 in at least one band, see Cutri et al. 2012 for additional details); rv = variable in the radio bands at 1.4 GHz. \\
Col. (7) Estimate level of confidence derived from the KDE analysis.\\
Col. (8) Redshift: ? = unknown. \\
Col. (9) Results of the comparison with previous analyses. 1 = UGS analyzed in Massaro et al. (2013a) , 2 = UGS analyzed in Massaro et al. (2013b)
3 = UGS analyzed in Paggi et al. (2013). Exclamation mark (!) indicates that the $\gamma$-ray blazar candidate is the same IR source found
in the previous investigation.
\label{tab:table2}
\end{table*}

We note that the comparison between the $\gamma$-ray blazar candidates found in the SU and in the SDA samples 
and those presented in Massaro et al. (2013b) based on the WENSS radio analysis was not possible because the
footprints of the surveys used did not overlap.
We also verified that the selected $\gamma$-ray blazar candidates having a SDSS counterpart
exhibit optical color consistent with those of BL Lacs \citep[i.e., $u-r<1.4$, see][for more details]{massaro12}.
We found that with the only exception of NVSSJ154824+145702 all of them 
have the same optical properties of the BZB population.

Within the whole sample of UGSs analyzed, 
there are 25 sources that were also unidentified in the 1FGL \citep{abdo10} and 
were analyzed on the basis of two different statistical approaches: the Classification Tree
and the Logistic regression analyses \citep[see][and references therein]{ackermann12}.
By comparing the results of our association method with those in Ackermann et al. (2012), we found that
19 out of 25 UGSs with a $\gamma$-ray blazar candidate recognized according to our method
are also classified as AGNs. All of them with a probability higher than 60\% with 14 higher than 80\%. 
The remaining three sources were classified as pulsar candidates but with a very low probability (i.e. $\leq$60\%)
Consequently, our results are in good agreement with the classification 
suggested previously by Ackermann et al. (2012) and thus consistent with the $\gamma$-ray AGN nature.

Finally, we remark that several $\gamma$-ray pulsars have been identified after the release of the 2FGL, where they are listed as UGSs.
However, we did not exclude these UGSs from our sample to test if, as expected, we did not find any blazar-like counterpart
associable to them. Thus, in agreement with our expectations, all the UGSs for which we found a $\gamma$-ray blazar candidates
do not have any pulsars associated according to the Public List of LAT-Detected Gamma-Ray Pulsars 
\footnote{\underline{https://confluence.slac.stanford.edu/display/GLAMCOG/Public\\+List+of+LAT-Detected+Gamma-Ray+Pulsars}}.

\section{Summary and conclusions}
\label{sec:summary}
In this paper we presented an non-parametric method to search for $\gamma$-ray blazar candiates
within two samples of UGSs. 
First we identify all the radio sources in the two major surveys \citep[i.e., NVSS and SUMSS][respectively]{condon98,mauch03}
that lie within the positional uncertainty region at 95\% level of confidence, then we investigate the IR colors
of their \wse\ counterparts to recognize those with similar spectral properties in the simple [3.4]-[4.6]-[12] color-color plot.
With respect to our previous \wse\ selection of $\gamma$-ray blazar candidates \citep[e.g.,][]{paper3,ugs1}
the criteria adopted in the present analysis are less conservative, since the detection of the \wse\ counterpart at 22$\mu$m is not required.
A small fraction ($\sim$8\%) of the \fer\ blazar are in fact not detected at 22$\mu$m. 
Thus, to compare the IR colors of the \fer\ blazars with those of the radio sources selected, we adopted a KDE technique 
as already presented in Massaro et al. (2011a), Massaro et al. (2012a) and more recently in Paggi et al. (2013).
Our new approach, being less restrictive than those adopted in our previous associations,
permits to search for faint $\gamma$-ray blazar candidates that were not previously selected because too faint at 22$\mu$m.
By relaxing the requirement on the detection at 22$\mu$m 
and thus on the [12]-[22] color, this method would select candidate blazars at the cost of a larger contamination, 
mitigated by the requirement on the presence of a radio counterpart.

We found 41 and 14 radio sources with IR similar to those of the \fer\ blazars within the NU and the SU samples, respectively.
In addition, we investigated the sample of radio sources discovered with recent deep ATCA observations performed
to search for radio counterparts of the UGS in the southern hemisphere. Among 416 radio objects listed in Petrov et al. (2013)
only 11 sources have an IR counterpart consistent with the $\gamma$-ray blazars.
The total number of $\gamma$-ray blazar candidates is 66 all listed in Table~\ref{tab:table1} and Table~\ref{tab:table2}.
without no multiple candidates within the positional uncertainty regions of the UGSs analyzed.
{ We estimate a probability of spurious association for the $\gamma$-ray blazar candidates selected according to our method
of the order of 4\% and 3\% for the NU and SU samples, respectively.}

It is worth noting that the large majority of our candidates show IR colors more consistent with the region occupied by the BZBs
in the [3.4]-[4.6]-[12] $\mu$m color-color diagram rather than that of BZQs.
Thus they could be potential faint and so distant BZBs that were not previously selected with different methods
because lacking of the IR flux at 22$\mu$m. 
More detailed investigations based on ground-based, optical and near IR, 
spectroscopic follow up observations will be planned for the selected $\gamma$-ray blazar candidates
to confirm their nature and to obtain their redshifts.
\begin{table}
\tiny
\caption{Optical magnitudes for the \wse\ counterparts.}
\begin{tabular}{|lcccccc|}
\hline
WISE      &  B1  &  R1  &  B2  &  R2  &  I   & $\theta$ \\
name      &  mag &  mag &  mag &  mag &  mag & arcsec      \\
\hline
\noalign{\smallskip}
  J003119.70+072453.6 & 19.03 & 18.17 & 19.84 & 18.63 & 18.67 & 0.14\\
  J003908.14+433014.6 & 19.9 & 19.61 & 21.42 & 20.77 &  & 0.14\\
  J010345.73+132345.4 & 17.98 & 17.73 & 18.69 & 17.38 & 17.24 & 0.07\\
  J011619.59-615343.5 &  & 17.72 & 18.22 & 17.78 & 17.91 & 0.27\\
  J013306.35-441421.3 &  & 18.38 & 19.7 & 18.12 & 18.76 & 0.26\\
  J014347.39-584551.3 &  & 16.7 & 18.48 & 16.64 & 17.04 & 0.04\\
  J015248.80+855703.6 & 20.57 & 18.84 & 19.63 & 18.71 & 17.82 & 0.38\\
  J020020.94-410935.6 &  & 19.84 & 21.1 & 18.79 & 18.75 & 0.6\\
  J022744.35+224834.3 &  &  & 20.82 & 20.22 & 19.28 & 0.35\\
  J031240.54+201142.8 &  & 19.34 & 21.22 & 19.42 & 19.07 & 2.63\\
  J031614.31-643731.4 &  & 16.59 & 18.19 & 16.57 & 16.82 & 0.22\\
  J033153.90+630814.1 &  &  & 20.66 & 19.92 & 18.35 & 0.35\\
  J034022.89-242407.2 &  & 19.56 & 20.07 &  &  & 0.21\\
  J035309.54+565430.8 & 20.09 & 19.24 & 20.43 & 18.76 & 18.53 & 0.55\\
  J040946.57-040003.4 & 19.45 & 19.18 & 17.53 & 16.98 & 16.86 & 0.07\\
  J041605.81-435514.6 &  & 18.49 & 18.7 & 18.17 & 18.0 & 0.18\\
  J042025.09-374445.0 &  & 20.44 & 20.73 & 19.71 & 18.17 & 0.38\\
  J052313.07-253154.4 &  & 19.2 & 20.83 & 20.07 & 18.95 & 0.17\\
  J055618.74-435146.1 &  & 19.23 & 18.88 & 19.08 & 18.08 & 0.31\\
  J060102.86+383829.2 &  & 19.11 &  & 19.84 & 18.48 & 0.04\\
  J064435.72+603851.2 & 20.01 & 19.58 & 20.7 & 18.75 & 18.37 & 0.3\\
  J065845.02+063711.5 & 20.25 &  &  & 19.12 & 18.3 & 0.39\\
  J072354.83+285929.9 & 19.78 & 19.05 & 19.97 & 18.72 &  & 0.19\\
  J074627.03-022549.3 & 19.03 &  & 18.59 & 18.43 & 16.53 & 0.31\\
  J092849.83-352948.9 &  & 18.56 & 19.64 & 18.07 & 18.23 & 0.23\\
  J093754.72-143350.3 & 18.82 & 17.92 & 18.64 & 17.73 & 17.56 & 0.1\\
  J101544.44+555100.7 & 19.69 & 19.42 & 20.61 & 19.35 &  & 0.37\\
  J103015.35-840308.7 &  & 19.36 & 19.26 & 18.84 & 18.03 & 0.15\\
  J111511.74-070239.9 &  & 19.86 & 20.68 & 19.05 & 18.66 & 0.14\\
  J112325.38-252857.0 & 16.9 & 15.76 & 15.87 & 15.56 & 15.51 & 0.19\\
  J112903.25+375657.4 & 19.9 & 19.23 & 19.35 & 19.48 & 18.58 & 0.65\\
  J122358.17+795327.8 &  & 17.6 & 20.18 & 18.46 & 17.63 & 1.04\\
  J125422.47-220413.6 &  & 19.88 & 18.67 & 19.11 & 18.22 & 0.41\\
  J125949.80-374858.1 &  & 17.44 & 18.07 & 16.78 & 17.35 & 0.17\\
  J131552.98-073301.9 & 19.78 & 18.68 & 18.75 & 17.75 & 17.56 & 0.16\\
  J132840.61-472749.2 &  & 17.75 & 18.23 & 16.8 &  & 0.98\\
  J133916.44-234829.4 & 20.3 & 19.3 & 20.43 & 19.79 & 18.5 & 0.31\\
  J134042.02-041006.8 & 18.21 & 17.21 & 17.59 & 16.46 & 17.08 & 0.19\\
  J134543.05-335643.3 &  & 17.98 & 19.58 & 18.65 & 18.12 & 0.38\\
  J134706.89-295842.3 & 17.85 & 17.09 & 18.8 & 17.14 & 17.09 & 0.41\\
  J151303.66-253925.9 & 19.92 & 18.96 & 19.77 & 20.35 &  & 0.5\\
  J151649.26+365022.9 & 20.9 &  & 21.49 & 20.07 & 19.16 & 1.58\\
  J154824.39+145702.8 & 20.51 & 18.29 & 19.86 & 17.74 & 17.45 & 0.41\\
  J164619.95+435631.0 & 20.43 & 19.73 & 20.42 & 19.67 &  & 0.34\\
  J170409.59+123421.7 & 19.86 & 18.04 & 18.62 & 17.63 & 17.46 & 0.47\\
  J170433.84-052840.6 & 19.62 & 18.97 & 18.42 & 17.28 & 17.98 & 0.45\\
  J200506.02+700439.3 & 20.73 & 19.25 & 19.24 & 18.65 &  & 0.45\\
  J202155.45+062913.7 & 17.27 & 16.13 & 17.01 & 16.67 & 16.03 & 0.43\\
  J203451.08-420038.2 &  & 18.97 & 19.34 & 18.87 & 18.27 & 0.44\\
  J204201.92-731913.5 &  & 17.46 & 17.9 & 18.36 & 18.04 & 0.29\\
  J211522.00+121802.8 & 18.15 & 18.15 & 17.68 & 17.31 & 17.58 & 0.16\\
  J213253.05+261143.8 & 20.04 & 19.29 & 19.14 & 19.62 & 18.44 & 0.07\\
  J213430.18-213032.6 & 19.77 & 18.65 & 18.96 & 16.8 & 17.7 & 0.09\\
  J213349.21+664704.3 &  &  &  & 19.37 & 18.8 & 0.45\\
  J221330.33-475425.0 &  & 18.12 & 18.6 & 18.34 & 18.33 & 0.05\\
  J222830.19-163642.8 & 18.57 & 19.34 & 19.95 & 19.04 & 17.91 & 0.29\\
  J224604.98+154435.3 & 19.14 & 18.27 & 19.57 & 18.53 & 17.65 & 0.13\\
  J225128.69-492910.6 &  & 18.8 & 19.21 & 18.45 & 18.03 & 0.42\\
  J234302.29-475749.9 &  & 19.84 & 18.92 & 21.3 & 18.32 & 0.29\\
  J235836.83-180717.3 & 19.14 & 18.45 & 18.28 & 17.22 & 17.53 & 0.3\\
\noalign{\smallskip}
\hline
\end{tabular}\\
\label{tab:optical}
\end{table}

\acknowledgements
{ We thank the anonymous referee for useful comments on the probability of spurious associations 
that improved our paper.}
The work is supported by the NASA grants NNX12AO97G.
R. D'Abrusco gratefully acknowledges the financial 
support of the US Virtual Astronomical Observatory, which is sponsored by the
National Science Foundation and the National Aeronautics and Space Administration.
The work by G. Tosti is supported by the ASI/INAF contract I/005/12/0.
Howard A. Smith acknowledges partial support from NASA-JPLRSA contract 717437.
TOPCAT\footnote{\underline{http://www.star.bris.ac.uk/$\sim$mbt/topcat/}} 
\citep{taylor05} for the preparation and manipulation of the tabular data and the images.
The WENSS project was a collaboration between the Netherlands Foundation 
for Research in Astronomy and the Leiden Observatory. 
We acknowledge the WENSS team consisted of Ger de Bruyn, Yuan Tang, 
Roeland Rengelink, George Miley, Huub Rottgering, Malcolm Bremer, 
Martin Bremer, Wim Brouw, Ernst Raimond and David Fullagar 
for the extensive work aimed at producing the WENSS catalog.
Part of this work is based on archival data, software or on-line services provided by the ASI Science Data Center.
This research has made use of data obtained from the High Energy Astrophysics Science Archive
Research Center (HEASARC) provided by NASA's Goddard
Space Flight Center; the SIMBAD database operated at CDS,
Strasbourg, France; the NASA/IPAC Extragalactic Database
(NED) operated by the Jet Propulsion Laboratory, California
Institute of Technology, under contract with the National Aeronautics and Space Administration.
Part of this work is based on the NVSS (NRAO VLA Sky Survey);
The National Radio Astronomy Observatory is operated by Associated Universities,
Inc., under contract with the National Science Foundation. 
This publication makes use of data products from the Two Micron All Sky Survey, 
which is a joint project of the University of 
Massachusetts and the Infrared Processing and Analysis Center/California Institute of Technology, 
funded by the National Aeronautics and Space Administration and the National Science Foundation.
This publication makes use of data products from the Wide-field Infrared Survey Explorer, 
which is a joint project of the University of California, Los Angeles, and 
the Jet Propulsion Laboratory/California Institute of Technology, 
funded by the National Aeronautics and Space Administration.

{}

\end{document}